\documentclass[twocolumn,preprintnumbers,amssymb,amsmath,9pt]{revtex4}[10pt]
\usepackage{graphicx}
\usepackage{dcolumn}
\usepackage{bm}
\usepackage{amssymb}
\usepackage{epsfig}
\newcommand{\be}{\begin{equation}}
\newcommand{\ee}{\end{equation}}
\newcommand\beq{\begin{eqnarray}}
\newcommand\eeq{\end{eqnarray}} 
\newcommand\eqn[1]{\label{eq:#1}} 
\newcommand\eq[1]{eq. (\ref{eq:#1})} 
\newcommand{\la}{\langle}
\newcommand{\ra}{\rangle}
\newcommand{\vev}[1]{\langle #1 \rangle}
\newcommand{\mpi}{m_\pi}
\newcommand{\fpi}{f_\pi}
\newcommand{\eV}{{\rm ~eV }}

\newcommand{\GeV}{{\rm ~GeV }}

\newcommand{\MeV}{{\rm ~MeV }}

\begin{document}

\preprint{INT-PUB 05-17}

\title{Exotic axions}
\author{David B. Kaplan}
\email{dbkaplan@phys.washington.edu}
\affiliation{Institute for Nuclear Theory, University of Washington, Seattle, WA 98195}
\author{Kathryn M. Zurek}
\email{zurkat@u.washington.edu}
\affiliation{Institute for Nuclear Theory, University of Washington,
  Seattle, WA 98195}

\begin{abstract}

We show that axion phenomenology may be significantly different than
conventionally assumed
 in theories which exhibit late phase transitions (below the
QCD scale).  In such theories one can find multiple pseudoscalars with
axion-like couplings to matter, including a string scale axion, whose
decay constant far exceeds the conventional cosmological bound.  Such theories have
several dark matter candidates.

\end{abstract}
\date{\today}
\maketitle

\section{Introduction}
The smallness of the experimentally determined upper bound on the strong CP violating parameter,
$\bar\theta\lesssim 10^{-9}$, is an outstanding puzzle of the standard
model.  One can either assume CP to be  an exact symmetry
spontaneously broken in such a way as to ensure that $\bar\theta$ is naturally small,
as in the Nelson-Barr mechanism \cite{Nelson:1983zb,Barr:1984fh}, or one
can introduce a  $U(1)$ symmetry, known as the Peccei-Quinn (PQ)
symmetry \cite{Peccei:1977hh,Peccei:1977ur} to allow  $\bar\theta$ to
dynamically relax to zero. (Models suppressing strong CP violation via spontaneously broken parity
symmetry have also been constructed \cite{Mohapatra:1978fy,Babu:2001se}.) An attractive feature of the PQ mechanism is that
it divorces the strong CP problem from flavor physics---the masses and
mixings of the quarks---whose origin
remains a mystery. The PQ mechanism entails a global $U(1)$ symmetry
which is exact up to a QCD (and possibly electromagnetic) anomaly.
The symmetry breaks spontaneously at a scale $f$, giving rise to a  pseudoscalar
Goldstone boson, the axion
, which couples to matter via the
interaction $ (a/f) G\tilde G$
\cite{Weinberg:1977ma,Wilczek:1977pj,Kim:1979if,Shifman:1979if,Zhitnitsky:1980tq,Dine:1981rt}.
Here $a$ is the axion, $f$ is its 
decay constant, $G_{\mu\nu}$ is the gluon field strength, and the ratio $(a/f)$ should
be thought of as an angle. This angle has a potential arising from
instantons which causes $(a/f)$ to sit at the vacuum where $\bar \theta=0$.  The axion
will in general have additional derivative couplings to matter, such
as a model dependent coupling to photons of the form $(a/f) F\tilde F$
\cite{Kaplan:1985dv,Srednicki:1985xd}.     The mass of the axion $m_a$ 
satisfies $m_a \approx  m_\pi f_\pi/f$ where $m_\pi\approx
140\MeV$ and
$f_\pi\approx 93\MeV$ are the pion mass and decay constant respectively.

The axion decay constant is bounded from below by  collider experiments and 
astrophysical arguments. The latter are the more stringent: if  $f$ is too small,
the coupling to ordinary matter  is large enough to allow
 copious axion production in red giants and supernovae, leading to an
 unacceptably rapid cooling rate. This yields the lower bound
 $10^9\GeV\lesssim f$ \cite{Raffelt:1996wa}.  It has also been shown that in
conventional axion models, large $f$ leads to copious production of
cold, degenerate axions in the early universe
\cite{Preskill:1982cy,Abbott:1982af,Dine:1982ah}, so that  $f\gtrsim
10^{12}\GeV$ leads to unacceptably 
large
$\Omega_{dm}$, where $\Omega_{dm}$ is the fraction of dark
matter in the universe today, measured to be $0.21$ to within
$4\%$ \cite{PDBook}. Therefore the axion decay constant
is conventionally assumed to lie in the window $10^9\GeV\lesssim
f\lesssim 10^{12}\GeV$.  At the upper end of this bound 
axions are a viable dark
matter candidate, and experimental attempts to detect
their presence are in progress \cite{Rosenberg:2004zc}.

In this Letter we will present a scenario for axions  which exhibits
exotic axion cosmology, and so we summarize here 
the conventional picture. The PQ symmetry breaks 
spontaneously  at a temperature $T\sim f$ well 
above the QCD scale where the instanton induced axion potential turns
on.  Following this phase transition  $(a/f) $ equals some random angle $\theta_i$ until $T\sim 
1\GeV$, when the axion potential develops rapidly.  The axion field
begins to oscillate at temperature $T_i\sim 1\GeV$ when the axion mass comes
within the horizon, $m_a(T_i)\approx H(T_i)$, where $H(T_i)$ is the Hubble
parameter at temperature $T_i$, which is  only weakly dependent on
$f$. The coherent oscillation may be thought of as a gas of 
degenerate nonrelativistic axions with  number density of axions per comoving
volume  equal to $n_a
= \theta_i^2 H(T_i) f^2$.  This quantity remains constant since
the axions are nonrelativistic and because their annihilation rates are
negligible. As a result, the subsequent axion energy density at temperature $T$
is given by $\rho_a=m_a n_a (R(T_i)/R(T))^3$, where $R(T)$ is the
Robertson-Walker scale factor at temperature $T$. The upper bound on
$f$ follows from observational limits on $\Omega_{dm}$ since $\rho_a$ varies almost linearly with $f$,
assuming that $\theta_i=O(1)$.

 There have been prior attempts to evade the
cosmological bound, motivated in part by the fact that in string
theories, axions with $f = \sqrt{2}\alpha_U M_p\approx10^{16}\GeV$,
where $\alpha_U$ is the unified value of the fine structure constant,
are generic. A 
trivial way to  harmlessly incorporate an axion $A$ with decay constant
$F>10^{12}\GeV$  is to introduce a
second pseudoscalar $a$ so that the Lagrangian contains the term
$(A/F + a/f)G\tilde G$. Provided that the simultaneous transformation $A\to (A+
\epsilon F)$,  $a\to (a - \epsilon f)$ is an exact symmetry
of the full theory, the spectrum will consist of a innocuous, massless
Goldstone boson plus an axion with decay constant $f$ subject to the
usual constraints summarized above.  Another resolution is to
assume inflation and an ensemble of initial angles $\theta_i$ to
choose from, invoking the anthropic principle to justify a value in
our universe of  $\theta_i\approx 0$ 
\cite{Linde:1987bx}; recently it was pointed out that such a scenario
could be constrained by the PLANCK
polarimetry experiment \cite{Fox:2004kb}. In addition, Banks, Dine and
Graesser have pointed out that any attempt to evade the cosmological
upper bound on the axion energy must deal with the generally more
difficult question of the cosmological saxion energy problem \cite{Banks:2002sd}.  Another
 interesting proposal by Barr and Kyae (BK) \cite{Barr:2004nj}
is an  axion coupled to the light quarks in such a way to allow for
exotic cosmological evolution of the axion below the QCD scale. Our
motivation here is to explore unconventional axion cosmology without
invoking the anthropic principle, and without convoluting the PQ mechanism with
flavor physics; in common with the BK suggestion, and the earlier 
 ref.~\cite{Barr:1999mk},  our scenario
involves late evolution of the PQ symmetry breaking order parameter,
although it differs significantly in realization and phenomenology.

The conventional  cosmological bound on $f$ results because the PQ
symmetry spontaneously breaks at a temperature well before the QCD
scale. However, if  $f$ only evolves to a large vev after
the QCD time, such bounds may  be  evaded.  This could be
accomplished with a sufficiently flat PQ potential, so that the radial
mode only evolves out to a large vev once its mass enters the horizon.

The fine 
tuning associated with such a flat potential can only be avoided with
supersymmetry (SUSY); but even then SUSY breaking will in general generate a
curvature for the PQ potential which forces PQ breaking at or above
the weak scale, a transition which is too early to evade the
cosmological constraints.
This problem can be avoided with the introduction of a new  sector
which couples only weakly to the standard model via the conventional
PQ sector.  We show that such a coupling
may be weak enough to shield the potential in the new sector from SUSY
breaking effects, 
while still significantly
affecting 
axion phenomenology.  The spectrum of this theory includes (i) an axion
far lighter than conventionally allowed with no significant
cosmological abundance; (ii) an additional  axion-like particle
which is heavier than would be an axion with comparable decay
constant; (iii) a light dilaton-like scalar particle.  The latter two
are dark matter candidates. In the next sections, we describe  a model
 which realizes this scenario.

\section{A  model}

Our starting point is to assume a viable supersymmetric theory
which implements the conventional PQ mechanism.  We
assume that there exists a  superfield $\phi_1$ 
which carries PQ charge and couples to  colored fermions in a real
representation of the gauge group; at a
temperature well above the QCD scale this field
acquires a vev $\vev{\phi_1}=v_1/\sqrt{2}$ which lies  within the 
conventional  window $10^9\GeV \lesssim v_1 \lesssim
10^{12}\GeV$. With this vev, the heavy colored fermions coupled to
$\phi_1$ develop a mass
 $M_Q = g v_1$.  
It is important for our modification of the theory that  that the saxion
be light (to be specified below) so that it not communicate
large SUSY breaking to a new sector we 
will be adding. A light saxion is expected in any theory of low
energy SUSY breaking; it could also occur in gravity mediated SUSY
breaking models so long as the PQ sector is sequestered from the SUSY
breaking. An excess of saxion energy can be avoided in such models
either by having relatively late inflation (below the PQ breaking
scale but well above the QCD time), or by having the minimum of the
saxion potential be at the same point as preferred by finite
temperature effects prior to an  epoch of higher scale inflation. 


To this theory we now introduce the superpotential
\be
\widetilde W=\sqrt{2}\lambda A (h \phi_1\phi_2-\phi_0^2)
+\frac{1}{\sqrt{2}}\mu^2 B\left(2(\phi_0/v_0)^2-1\right)\ ,
\eqn{w}
\ee
where $\phi_2$ carries opposite PQ charge from $\phi_1$, while $A$,$B$ and
$\phi_0$ are PQ-invariant fields. 
 The
four
parameters  $v_0^2$, $\mu^2$, $h$ and $\lambda$ may all be taken to be
real by redefinition of the phases  of $A$, $B$, $\phi_0$ and $\phi_2$.  The
part of the scalar potential  relevant to 
us is
\begin{eqnarray}
\widetilde V(\phi_1,\phi_2,\phi_0) =2\lambda^2|h\phi_1\phi_2-\phi_0^2|^2
+\frac{\mu^4}{2} |2(\phi_0/v_0)^2-1|^2
\end{eqnarray}
The first term in
$\widetilde V$ exhibits a flat direction in $\phi_2$ and $\phi_0$, which is
lifted slightly by the second term in $\widetilde V$, with $\lambda\ll 1$
and  $\mu\ll \Lambda_{QCD}$.  The smallness of the couplings leave
$\sqrt{2}\vev{\phi_1}=v_1$  unaffected, and the  minimum of the almost flat
direction is at
$\sqrt{2}\vev{\phi_0}=v_0\gg v_1$, and
$\sqrt{2}\vev{\phi_2}=v_2=v_0^2/(h v_1)\gg 
v_0$.  

We assume that following inflation and
reheating the Universe sits away from the minimum of the potential with
 $\sqrt{2}\vev{\phi_1}=v_1$, and $\vev{A}=\vev{B}=\vev{\phi_0}=\vev{\phi_2}=0$,
the latter being determined by high temperature effects due to
interactions with unspecified heavy fields  prior to
inflation. This field configuration persists down to a temperature
$T_0<\Lambda_{QCD}$, satisfying $\mu^2/v_0 
\approx H(T_0)$, when the curvature lifting the flat direction is
sufficiently strong to overcome the Hubble friction.   Then $\phi_0$ will
roll out to its minimum at $\la \phi_0\ra=v_0\gg v_1$, 
causing $\phi_2$ to follow its flat direction out to $v_2=v_0^2/(h
v_1)\gg v_0$.  With $\mu \ll \Lambda_{QCD}$ this phase transition will
occur at
$T_0<\Lambda_{QCD}$, and as we discuss further below,  the dark matter
produced in this transition can be made acceptable.

To understand how the pseudoscalars behave in
this model, we expand around the minimum of the potential, writing
$
\phi_i = \frac{1}{\sqrt{2}}(v_i+\sigma_i) e^{i \pi_i/v_1}$ with $
i=0,1,2$. The fields $\sigma_i$ and $\pi_i$ are the scalar and
pseudoscalar excitations respectively.
After adding  to $\widetilde V$ the QCD contribution to the $\pi_1$
potential, the complete pseudoscalar
potential is
\beq
V_\pi&=&-\mpi^2\fpi^2\cos{\frac{\pi_1}{v_1}} -\lambda^2  v_0^4
\cos\left(\frac{\pi_1}{v_1}+\frac{\pi_2}{v_2}-\frac{2\pi_0}{v_0}\right)\cr 
&& -\mu^4\cos\left({\frac{2\pi_0}{v_0}}\right)
\eeq
 (up
to an overall O(1) factor \cite{Kaplan:1985dv} in front of the first term
 which does not concern us here). 
Since $\pi_1$ is the only field that couples to ordinary matter, to
understand phenomenology we must decompose $\pi_1$
into mass eigenstates $a_{1,2,3}$.  We diagonalize the mass matrix to leading nonzero
order in $\delta= v_1/v_0$ and $\epsilon=\mu^4/(\lambda^2v_0^4)$, and to all
orders in $x=\lambda^2 v_0^4/(m_\pi^2 f_\pi^2)$, finding the masses
and decay constants
\beq
\begin{aligned}
m_{a_1}^2 & =  \frac{\mpi^2 \fpi^2}{ v_1^2}\left(1+x\right)\ , \ \ 
&f_1=&v_1\ ,\\
m_{a_2}^2 & =  \frac{m_\pi^2 f_\pi^2}{v_0^2}\left(\frac{4x}{1+x}\right) \ , 
&f_2=&v_0\left(\frac{1+x}{2x}\right)\ ,\\
m_{a_3}^2 & =  \frac{\mu^4}{v_2^2} \ ,&
\eqn{axionmass}
\end{aligned}
\eeq
where the decay constants are defined by
\beq
{\cal
  L}_{QCD}=\frac{\pi_1}{v_1}G\widetilde{G}= \left(\frac{a_1}{f_1}+\frac{a_2}{f_2}
\right) G\widetilde{G}\ .
\eqn{axionf}
\end{eqnarray}
In these formulas we have neglected terms of order $v_0/v_2$; the
$a_3$ pseudoscalar  decouples from the standard model to 
the order we work.

This model yields a distinct particle spectrum:
$a_1$ is mostly $\pi_1$ and  couples like an axion with decay constant $f_1=v_1$, except
that its mass can be much heavier than that of a conventional axion by the
factor $\sqrt{1+x}$;  $a_2$ looks like a very light axion,  primarily composed of $\pi_0$, with
decay constant
$f_2= v_0(1+x)/2x\approx v_0/2$.  Since $v_0 \gg v_1$,  $m_2$ is much
lighter than that allowed by the conventional axion bounds --- it is
potentially an axion with string scale PQ constant $f_2\approx
10^{16}\mbox{ GeV}$. 
The energy
originally in $\pi_1$ is primarily 
transferred into the $a_1$ field, making it a good dark matter
candidate;   the light  $a_2$ axion
receives only a small fraction of that energy, suppressed by $(v_1/v_0)^2\ll
1$.

\section{ Constraints}

At temperatures below the QCD scale but above the secondary transition, the universe
has a background density of cold $\pi_1$ pseudoscalars which behave like a
conventional axion with decay constant $v_1$.  Subsequently the $\phi_0$
field rolls out to its minimum $\vev{\phi_0}=v_0$ at the
temperature $T_0$ where  $H(T_0)\sim \mu^2/v_0$. At this point (i)
the energy in $\pi_1$ pseudoscalars gets redistributed among the $a_1$ and
$a_2$ mass eigenstates following \eq{axionmass}; (ii) an energy density
$\mu^4$ is released and is mostly transferred into $\sigma_0$ (radial) oscillations
of the $\phi_0$ field. There are two cosmological constraints on the late transition.
The first is that  the combined
energy  density in the  exotic ${a_1}$ axion and  $\sigma_0$ scalar  not exceed the observed dark
matter density today.  A second constraint is  that the phase transition occur after the QCD time but
before today (or well before matter-radiation equality if either
${a_1}$ or $\sigma_0$ contribute an appreciable fraction of the dark
matter).     
In addition there is the astrophysical constraint that to prevent
copious production of any of the light pseudoscalars
in supernovae, we require $f_1 \mbox{, }f_2 \gtrsim 10^9\mbox{ GeV}$.
We now discuss these constraints in detail, and map out the
corresponding parameter space. 

\begin{figure}[t]
\includegraphics[width=3.0in.]{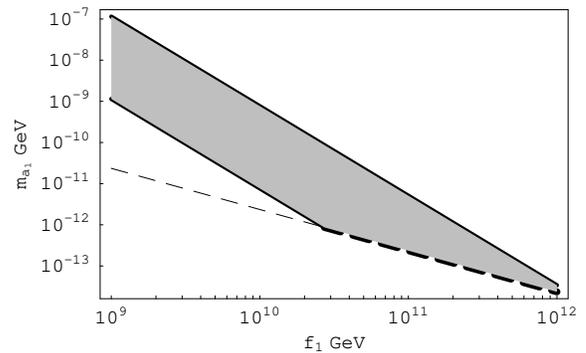}
\caption{The mass of the exotic axion $a_1$ vs. its decay constant $f_1$.
  Two solid parallel lines give
  $m_{a_1}(f_1)$ assuming $a_1$ 
  comprises half the dark matter for initial misalignment angle $\theta_i=0.1$ (upper line) and
  $\theta_i=1$ (lower line).  Dashed line is the axion mass in
  standard PQ models, with the heavy dashed line giving the
  region where conventional axions could be the dark matter.
 } 
\label{ma1f1}
\end{figure}

In order to ensure no excess of dark matter, we must limit the energy
in $\pi_1$ by requiring  $f_1\lesssim
10^{12}\GeV$.
We also require that the energy $\rho_0$ in the form of $\sigma_0$
oscillations produced at the secondary transition not dominate the
universe at the epoch of matter-radiation equality, $T_{eq}\approx1\eV$:
\be
\rho_0(T_{eq})\approx\mu^4\left(\frac{T_{eq}}{T_0}\right)^3\lesssim
T_{eq}^4\ ,\quad H(T_0)\approx \frac{\mu^2}{v_0}\ .
\eqn{Troll}
\ee 
A complication arises from the fact that energy in the early $\pi_1$
oscillations is transferred primarily into $a_1$ oscillations,
while $a_1$ becomes heavier than $\pi_1$ by a factor of
$\sqrt{1+x}$ as $\phi_0$ rolls from $\phi_0=0$ to $\phi_0=v_0$. This increase of energy
must come from the energy released during
the secondary transition; it can be represented  by a contribution to
the 
potential for $\phi_0$ due to the  $\phi_0$-dependent energy density of the
${a_1}$ axion condensate \footnote{ This is the same effect recently exploited  in the theory
of mass varying neutrinos (MaVaNs) \cite{Fardon:2003eh}}, of the form 
\begin{eqnarray}
\rho_1(\phi_0,T)&=& n_{\pi_1}(T)\, m_{\pi_1}
\sqrt{1+x(\phi_0)}\ ,
\label{rhob}
\end{eqnarray}
where $T_i\sim 1\GeV$ is the temperature when $\pi_1$ begins to
oscillate, $n_{\pi_1}(T)=\left(\theta_i^2
  H_if_1^2\right)(T/T_i)^3$ is the  number density of $\pi_1$ bosons
in the  temperature range $\Lambda_{QCD}\ge T\ge T_0$,
$m_{\pi_1}=(m_\pi f_\pi/f_1)$ is the $\pi_1$ mass before $\phi_0$ rolls out,  and
$x(\phi_0)= \left(\phi_0/v_0\right)^4x$  controls how the $\pi_1$ mass becomes the
$a_1$ mass as $\phi_0$ increases.
If $\rho_1(v_0,T_0)>\mu^4$, this potential delays the secondary
transition, so that $\phi_0$ only gains its vev at some lower temperature, 
$T=T_0'$ satisfying  $\rho_1(v_0,T_0')\approx \rho_0(T_0')\approx
\mu^4$. After this delayed transition both the energy density in $\phi_0$ and $a_1$
oscillations remain comparable, diluting $\propto T^{-3}$.  After a
little algebra one finds that \eq{Troll} still holds in this case, but
is augmented by the constraint that there not be too much energy in
the cosmological $a_1$ abundance, 
\beq
T_{eq}^4 \gtrsim \left(\frac{T_{eq}}{T_0}\right)^3 n_{\pi_1}(T_0)\, m_{a_1} \ ,
\eqn{dmconstraint}
\eeq
with $m_{a_1}$ defined in \eq{axionmass} and $T_0\approx\sqrt{M_p
  \mu^2/v_0}$ from \eq{Troll}. 
If we now assume
$\rho_{1}+\rho_0=\rho_{DM}$, that is that $a_1$
and $\sigma_0$ together compose the dark matter, we can use 
\eq{dmconstraint} to compute allowed masses and couplings, $f_1$,
$m_{a_1}$, for the observable dark matter candidate $a_1$.  The result is
shown in fig.~1; evidently the  mass  and coupling of $a_1$ can differ
significantly from those of the conventional axion.

Finally, we obtain a constraint from requiring $\phi_0$ to roll after the
QCD phase transition, but somewhat before matter-radiation equilibrium if
either  ${a_1}$ or  $\sigma_0$ are to be the dark matter:
\be
T_{eq}\left(\frac{v_0}{M_{pl}}\right)^{1/2} \lesssim \mu
\lesssim T_{QCD}\left(\frac{v_0}{M_{pl}}\right)^{1/2}. 
\eqn{TrollS}
\ee
If the origin of dark matter lies elsewhere, then $T_{eq}$ in the
above equation is replaced by today's temperature.

The model as it stands possesses an exact discrete symmetry $S\to -S$
which is spontaneously broken and leads to domain walls.  This can be
avoided by breaking the symmetry explicitly, either with a small
linear term in the low energy superpotential  \eq{w}, or in the high energy interactions
such that  $\vev{S}\ne 0$ (but $\lesssim v_1$) after post-inflationary
reheating.

Aside from cosmological constraints, there is also a naturalness constraint.
Having a flat potential is critical for the late phase transition, so
we require that SUSY breaking terms induced by the
interactions between $\phi_1$, $\phi_2$ and $\phi_0$ do not lift the flat
direction too much.  The most important term is the soft mass
generated for $\phi_2$ from the SUSY breaking $\phi_1$ mass, for which
we require 
\be
\Delta m_2^2 f_2^2\approx \frac{m_{a_1}^2 m_{\phi_1}^2}{16 \pi^2} \lesssim \mu^4.
\eqn{softm2}
\ee
This imposes a significant new constraint on $\mu$. In addition, we
are assuming this sector is sequestered from gravity-mediated SUSY
breaking \cite{Randall:1998uk}.

\setlength{\extrarowheight}{4pt}
\begin{table}[t]
\begin{tabular}{c|lllllll}
& $\, \ m_{a_1}\ $&$\, \ f_1\ $&$\, \  m_{a_2}\ $&$\, \ f_2\ $&$\, \ x\ $&$\, \ \mu\ $&$ \ T_0(T_0')\ $ \\ \hline
I & $\ 10^{-2}$&$\ 10^{10}$&$\ 10^{-5}$&$\ 10^{15}$&\ 10&$\ 10^{-3}$&$\ 10^{-2}$  \\ \hline
II & $\ 1$&$\ 10^{10}$&$\ 10^{-9}$&$\ 10^{16}$&$\ 10^3$&$\ 10^{-6}$&$\ 10^{-6}$ 
\label{params}
\end{tabular}
\caption{Two sets of parameters (I,II) allowed by the constraints
  eqs.~(\ref{eq:Troll},\ref{eq:dmconstraint}-\ref{eq:softm2}).  Set I
  gives parameters for an prompt transition at temperature $T_0$,
  set II gives parameters for a delayed transition at $T_0'$.  All
  parameters are in GeV except the axion masses which are in eV.} 
\end{table}

The combination of constraints
eqs.~(\ref{eq:Troll},\ref{eq:dmconstraint}-\ref{eq:softm2}) yields a parameter space
too large to explore in detail here \footnote{There are no interesting
constraints on this model arising from searches for exotic
long-distance forces due to light scalar mesons with   couplings to the gluon field strength
$G_{\mu\nu}G^{\mu\nu}$.}. We give here instead two representative
sets of values in Table~I satisfying the constraints.

\section{Phenomenology}

Conventional axion models have fairly circumscribed phenomenology; if
one assumes that the dark matter consists of a conventional axion, then
the mass and coupling of the axion are related in a direct way, and
both lie in a fairly model independent range about $m_a\sim 10^{-5}\eV$,
$f_a\sim 10^{12}\GeV$ determined by the initial
axion misalignment $\theta_i$.  If the axion is not the dark matter,
its  mass can be heavier.  

In contrast we have shown how these simple relations can be greatly
modified in a theory with a late phase transition below the QCD
scale, as shown in table~I.  Dark matter in
this theory can be comprised of roughly 
equal parts of remnant pseudoscalar and scalar particles.

It will be interesting to examine whether the late phase transition
discussed in this Letter
could give rise to identifiable signatures, due to large isocurvature
fluctuations seeded at the time of the transition, over angular scales
which are small relative to the measured microwave
background fluctuations.

We are grateful for numerous conversations about this work with
C. Hogan and A. Nelson.  This work was
 supported by DOE grant DE-FGO3-00ER41132.

\bibliography{axion}
\bibliographystyle{apsrev}

\end{document}